\documentclass[twocolumn,aps,showpacs,prb,tightenlines,amsmath,amssymb,superscriptaddress]{revtex4}
\usepackage{graphicx}
\usepackage{amssymb}
\usepackage{txfonts}
\usepackage{dcolumn}
\usepackage{amsmath}
\usepackage{bm}

\usepackage{colordvi}

\begin{document}
\title{Effect of electron-electron scattering on spin dephasing in
a high-mobility low-density two dimensional electron gas}
\author{X. Z. Ruan}
\author{H. H. Luo}
\author{Yang Ji}
\thanks{Author to whom all correspondence should be addressed}
\email{jiyang@red.semi.ac.cn}
\author{Z. Y. Xu}
\affiliation{SKLSM, Institute of Semiconductors, Chinese Academy of
Sciences, Beijing, 100083, China}
\author{V. Umansky}
\affiliation{Braun Center for Submicron Research, Department of
Condensed Matter Physics Weizmann Institute of Science, Rehovot,
76100, Israel}
\date{\today}


\begin{abstract}
Utilizing time-resolved Kerr rotation techniques, we have
investigated the spin dynamics of a high mobility, low density two
dimensional electron gas in a GaAs/Al$_{0.35}$Ga$_{0.65}$As
heterostructure in dependence on temperature from 1.5\ K to 30\ K.
It is found that the spin relaxation/dephasing time under a magnetic
field of 0.5 T exhibits a maximum of 3.12\ ns around 14\ K,
superimposed on an increasing background with rising temperature.
The appearance of the maximum is ascribed to that at the temperature
where the crossover from the degenerate to the nondegenerate regime
takes place, electron-electron Coulomb scattering becomes strongest,
and thus inhomogeneous precession broadening due to D'yakonov-Perel'
(DP) mechanism becomes weakest. These results agree with the recent
theoretical predictions [Zhou {\em et al.}, PRB {\bf 75}, 045305
(2007)], verifying the importance of electron-electron Coulomb
scattering to electron spin relaxation/dephasing.
\end{abstract}

 \maketitle

In recent years, spin dynamics in semiconductors has attracted
considerable attention because of its potential application in the
spin-based devices.\cite{Awschalom:2002} The operation of these
devices requires spin lifetime long enough to achieve storage,
transport and processing of information. Therefore, a comprehensive
understanding of spin relaxation mechanism is a key factor for the
realization of these devices. It is generally accepted that the
D'yakonov-Perel' (DP) mechanism is the leading spin
relaxation/dephasing (R/D) mechanism in $n$-type zinc-blende
semiconductors.\cite{DP} This is caused by an wavevector ${\bf
k}$-dependent effective magnetic field ${\bf \Omega}({\bf k})$ from
the bulk inversion asymmetry,\cite{Dresselhaus} i.e., the
Dresselhaus term, and/or the structure inversion
asymmetry,\cite{Rashba} i.e., the Rashba term. The spin relaxation
rate can be determined by $\tau^{-1}=\langle{\bf \Omega}({\bf
k})^2\rangle\tau_P({\bf k})$, where $\tau_P({\bf k})$ is the
momentum relaxation time.\cite{Meier} As the electron-electron
Coulomb scattering does not contribute to the momentum relaxation
time $\tau_p$, it has long been widely believed that the
electron-electron Coulomb scattering is irrelevant in the spin
relaxation.\cite{Meier,Lau:2005,Lau:2001,Song,Averkiev,Krishnamurthy,Bleibaum}
However, it was first pointed out by Wu and Ning\cite{Wu:2000} that
in the presence of inhomogeneous broadening, any scattering,
including the spin conserving electron-electron Coulomb scattering,
can cause an irreversible spin relaxation and dephasing. This
inhomogeneous broadening can be the energy-dependent
$g$-factor,\cite{Wu:2000} the DP term,\cite{Wu:2001,Wu:2003} and
even the ${\bf k}$-dependent spin diffusion along a spacial
gradient.\cite{weng} In $n$-type GaAs quantum well, the importance
of the electron-electron scattering to the spin relaxation was
proved by Glazov and Ivchenko\cite{Glazov:2002} by using
perturbation theory and Weng and Wu\cite{Wu:2003} from a fully
microscopic many-body approach. In a temperature-dependent
experimental study of the spin relaxation in $n$-type (001) quantum
wells, Harley{\it et al.} indirectly verified the effects of the
electron-electron scattering on spin
relaxation.\cite{harley,Leyland} Nevertheless, the importance of the
Coulomb scattering to the spin relaxation/dephasing (R/D) has not
yet been widely accepted. Recently, Bronold {\it et
al.}\cite{Bronold} and Zhou {\it et al.}\cite{Wu:2007} predicted
that electron-electron scattering could lead to a maximum in the
spin R/D time as a function of temperature at the temperature where
the transition from the degenerate to the nondegenerate regime
occurs. The latter particularly pointed out that this maximum is
{\em solely} from the electron-electron Coulomb scattering in
samples with low electron density but high mobility, since in such
samples the electron-impurity scattering and the electron-ac-phonon
scattering could be effectively excluded at low temperature. An
experimental observation of such a maximum helps to nail down the
importance of the Coulomb scattering to the spin R/D.

 In this paper, we report on time-resolved measurements on such
kind of high mobility two dimensional electron gas (2DEG) with low
electron density in the low temperature regime from 1.5\ K to 30\ K.
With minimal excitation density, spin-polarized electrons are
injected and probed near the Fermi energy. The ensemble spin
dephasing time $T^\ast_2$ is measured via time-resolved pump-probe
Kerr rotation (TRKR). We find that the spin R/D time under a
magnetic field of 0.5\ T indeed exhibits a maximum of 3.12\ ns
around 14\ K and a monotonic increase background from 1.03\ ns  at
1.5\ K to 2.67\ ns at 30\ K. These features agree with the recent
theoretical predictions,\cite{Wu:2003,Bronold,Wu:2007}
demonstrating the importance of the electron-electron Coulomb
scattering to electron spin R/D in a high-mobility low-density 2DEG.

\begin{figure}[thb]
\includegraphics[width=0.9\columnwidth]{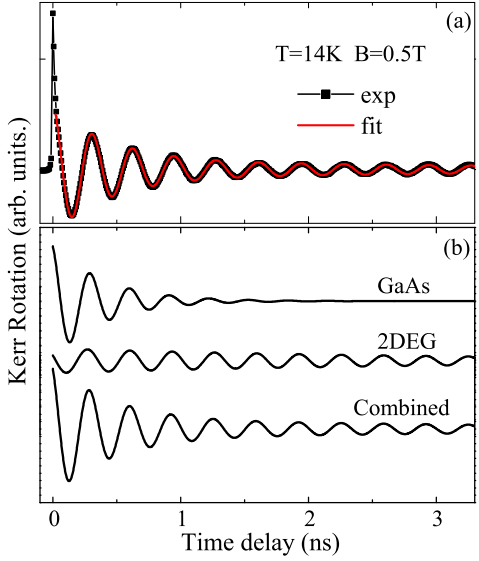}
\caption[FIG. 1]{(Color online) (a) Experimental TRKR trace (curve
with squares) at $T=14$\ K and $B=0.5$\ T. The solid line is the
fitting result. (b) Extracted TRKR signals of GaAs (top), 2DEG
(middle), and their combined TRKR signal (bottom).}
\end{figure}

The 2DEG sample used in our investigation contains a GaAs/AlGaAs
heterostructure grown by molecular beam epitaxy on a (001)-oriented
semi-insulating substrate. A 1400\ nm GaAs buffer layer was first
grown on the substrate followed by a 90\ nm undoped
Al$_{0.35}$Ga$_{0.65}$As spacer layer, 14\ nm $n$-doped
($3.1\times10^{18}$\ cm$^{-3}$) Al$_{0.35}$Ga$_{0.65}$As, a 10\ nm
undoped AlGaAs barrier layer, and finally a 7\ nm GaAs cap layer.
The 2DEG sample has a mobility of $3.2\times10^6$\
cm$^2$V$^{-1}$s$^{-1}$ and a density of $9.6\times10^{10}$\
cm$^{-2}$ at 4.2\ K. The TRKR measurements were performed in a
magneto-optical cryostat with a superconducting split-coil magnet.
The sample was excited near normal incidence with degenerate pump
and probe beams from a Ti:sapphire laser (76\ MHz repetition rate ).
The laser pulse has a temporal duration of $\sim$3\ ps and a
spectral width of $\sim$0.5\ meV, which allows for a high energy
resolution. The photon energy was tuned slightly above the band gap
of GaAs for the maximum Kerr rotation signal. The pump and probe
beams were focused to a spot of $\sim$100\ $\mu$m in diameter, with
constant powers of 200\ $\mu$W and 20 $\mu$W, respectively. The
circular polarization of the pump beam was modulated with
photoelastic modulator at 50 kHz for lock-in detection. The
circularly polarized pump beam incident normal to the sample surface
generated spin-polarized electrons with the spin vector along the
growth direction of the sample. The Kerr rotation $\theta(\Delta t)$
of a linearly polarized pulse after a time delay $\Delta t$,
measures the projection of the net spin magnetization as it
precesses about a magnetic field applied parallel to the sample
surface (in Voigt geometry).

\begin{figure}[thb]
\includegraphics[width=\columnwidth]{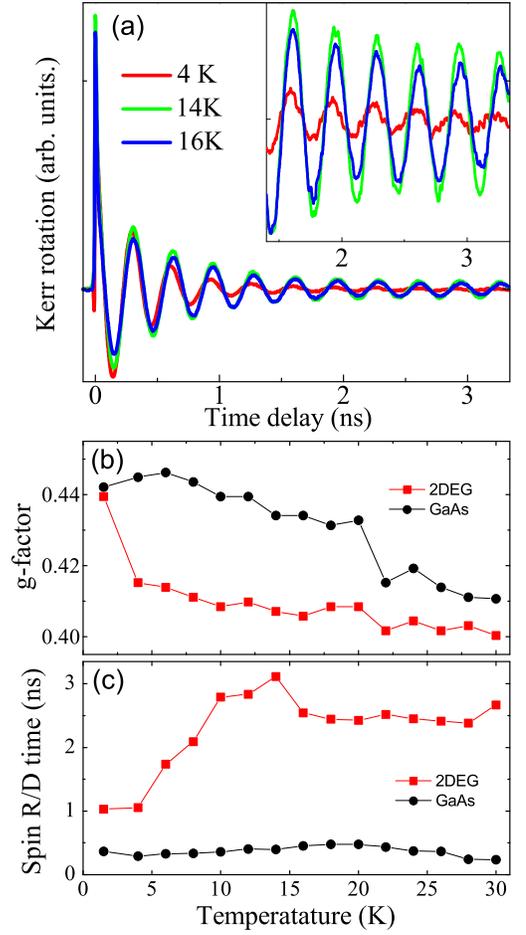}
\caption[FIG. 2]{(Color online) (a) TRKR traces at different
temperatures of 4 K (red), 14 K (green) and 16 K (blue). Inset:
zoomed picture of the same curve for the time delays between 1.4\ ns
and 3.34\ ns. (b) Electron $g$-factor as a function of temperature
for 2DEG (squares) and GaAs (circles). (C) Electron spin R/D time as
a function of temperature for 2DEG (squares) and GaAs (circles). All
data were taken at $B=0.5$ T and powers of pump : probe = 200\
$\mu$W : 20\ $\mu$W.}
\end{figure}

 A typical experimental TRKR trace measured at $T=14$\ K and $B=0.5$\ T is
presented in Fig. 1(a). The trace shows strong oscillations whose
frequency, i.e., the Larmor precession frequency $\omega$ gives the
electron $g$-factor by $\omega=g\mu_BB/\hbar$, where $\mu_B$ is the
Bohr magneton, $B$ is the transverse magnetic field, and $\hbar$ is
the reduced Planck's constant. The exponentially-decayed envelope
reflects the ensemble spin R/D time $T^\ast_2$. Quantitative
analysis shows that the experimental TRKR trace in Fig. 1(a)
contains oscillations with two different frequencies, rather than a
single frequency. This can be understood as follows. The photon
energy of pump and probe beams is only a little higher than the band
gap of GaAs. The 2DEG and the GaAs buffer layer are unavoidable to
be excited simultaneously. Spin-polarized electrons in the 2DEG and
the GaAs buffer layer both contribute to the Kerr rotation signal
with distinct precession frequencies. Therefore, the TRKR trace
shows two distinct precession frequencies (or $g$-factors). We can
extract the Kerr signal arising from the 2DEG or the GaAs buffer
layer through their distinct electron $g$-factors. The Kerr rotation
signal $\theta_K(\Delta t)$ as a function of time delay $\Delta t$
can be expressed as a superposition form of exponentially-decayed
harmonic functions for 2DEG and GaAs:
\begin{eqnarray}
\nonumber
 \theta_K(\Delta t)&=&A_1\exp(-\frac{\Delta
   t}{T^\ast_{21}})\cos(\omega_1 \Delta t+\phi_1)\\
& &\mbox{}+A_2\exp(-\frac{\Delta t}{T^\ast_{22}})\cos(\omega_2
                   \Delta t+\phi_2),
\end{eqnarray}
Where $A_1$ is the initial magnitude of electron spin polarization
in 2DEG, $T^\ast_{21}$ is the spin R/D time in 2DEG, $\omega_{1}$ is
the Larmor precession frequency in 2DEG, and $\phi_1$ is a phase
offset. $A_2$, $T^\ast_{22}$, $\omega_2$, and $\phi_2$ are the
corresponding parameters of GaAs.

Fitting the experimental data with Eq.\ (1) yields the solid curve
in Fig. 1(a). It is clearly seen that the fitting curve agrees very
well with the experimental data. A decomposition of the KR signal is
shown in Fig. 1(b). The decomposition uses the parameters obtained
from the fitting results in Fig. 1(a). The TRKR signal of 2DEG
indicates an electron spin R/D time of 3.12\ ns and an electron
$g$-factor of 0.407, while the TRKR signal of GaAs indicates an
electron spin R/D time of 0.40\ ns and an electron $g$-factor of
0.434. The combined signal of 2DEG and GaAs gives the fitting curve
in Fig. 1(a). Note that a very fast decay of the TRKR signal within
the first few picoseconds. We attribute this to the spin relaxation
of the photoexcited holes, which lose their initial spin orientation
very fast.\cite{Crooker:1997} Here we don't consider this fast
decay, i.e., hole spin relaxation.

Figure 2(a) shows TRKR traces under a magnetic field of 0.5\ T at
different temperatures of 4\ K, 14\ K, and 16\ K. One can find that
the oscillatory envelope decay becomes much slower from 4 K to 14 K,
and a little faster from 14 K to 16 K [see the inset of Fig. 2(a)].
These clearly indicate that the spin R/D time exhibits a maximum
around 14\ K. As the temperature was increased, we tuned the photon
energy of the pump and probe beams slightly above the band gap of
GaAs for the maximum Kerr rotation signal at a fixed time delay of
12 ps. Figure 2(b) displays the electron $g$-factors in 2DEG and
GaAs as a function of temperature from 1.5\ K to 30\ K. One can
clearly see that electron $g$-factor in GaAs at low temperatures is
about 0.44, which is a commonly accepted value in
GaAs.\cite{Kikkawa,Oestreich,Hohage} The electron $g$-factor in 2DEG
is smaller than that in GaAs. This is because the wavefunction of
electrons in the triangle quantum well penetrates into the potential
barrier AlGaAs. Except for the temperature of 1.5\ K, the electron
$g$-factors in 2DEG and GaAs are clearly resolved. From the distinct
$g$-factors in 2DEG and GaAs, we can obtain the corresponding
electron R/D time in 2DEG and GaAs. Figure 2(c) shows the
temperature dependence of electron R/D time in 2DEG and GaAs from
1.5\ K to 30\ K. A maximum of 3.12\ ns is clearly seen around 14\ K
in the electron spin R/D time of 2DEG as a function of temperature.
The maximum is superimposed on an increasing spin R/D time
background from 1.03\ ns at 1.5\ K to 2.67\ ns at 30\ K. The
electron spin R/D time in GaAs at different temperatures is around
0.4\ ns with moderate fluctuation. Similar temperature dependence of
the electron spin R/D time in bulk GaAs has been observed by the
previous work at low temperatures.\cite{Hohage}

The 2DEG sample used here is of high mobility, and thus the
electron-impurity scattering is weak. In addition, the
electron-impurity scattering has a very weak temperature dependence.
At very low temperature, the electron-ac-phonon scattering is
negligible.\cite{Yu} Therefore, the appearance of the maximum in the
spin R/D time as a function of temperature in Fig. 2(c) originates
from the electron-electron Coulomb scattering which dominates the
scattering process at low temperature. It is understood that
electron-electron Coulomb scattering has a nonmonotonic dependence
on temperature: at low temperature (degenerate limit), the
electron-electron scattering time $\tau_{ee}\propto T^{-2}$, while
at high temperature (nondegenerate limit), $\tau_{ee}\propto
T$.\cite{abrikosov,gio} The minimum of $\tau_{ee}$ appears at the
transition temperature where the crossover from the degenerate to
the nondegenerate regime occurs. Therefore, the contribution of
electron-electron Coulomb scattering to inhomogeneous precession
broadening due to DP mechanism has a minimum at the transition
temperature. Consequently, the spin R/D time versus temperature
curve exhibits a maximum. This feature agrees with the recent
theoretical prediction.\cite{Bronold,Wu:2007} Note that the Fermi
temperature ($T_F$) of the 2DEG estimated from the electron density
is about 40\ K, while the transition temperature is around 14\ K.
This deviation can be attributed to that the Fermi-Dirac
distribution function is strongly affected by the electron-electron
scattering in the intermediate temperature regime $T\sim
0.5T_F$.\cite{Hu,Flensberg} Thus, the transition temperature between
the degenerate and the nondegenerate regime in the 2DEG investigated
here is close to $0.5T_F$.

We now turn to discuss the increasing spin R/D time background with
rising temperature. For a low initial spin polarization, a large
increase of the spin R/D time with rising temperature has already
been observed by Brand {\it et al.}\cite{harley} and Stich {\it et
al.}.\cite{Stich:PRB} This behavior has been discussed from kinetic
spin Bloch approach by Weng and Wu\cite{Wu:2003} in  high
temperature regime and by Zhou {\it et al.}\cite{Wu:2007} in low
temperature regime. With both the experiment and calculation, Stich
{\it et al.} \cite{Stich:PRB} show that the spin R/D time increases
with rising temperature for low initial spin polarization in low
temperature regime, except that the spin R/D time peak was not
observed in their case. Using the method in Ref.\
\onlinecite{Stich:PRL} and taking into account the absorption ratio
between 2DEG and GaAs in this measurement, we estimate an initial
spin polarization degree of about 0.8\ \%. For such low initial spin
polarization, the inhomogeneous broadening determined by momentum
scattering in DP mechanism plays a dominant role.\cite{Stich:PRB} An
increasing temperature led to stronger momentum scattering, in other
words, a shorter momentum scattering time $\tau_P$. This in turn
induced an increasing spin R/D time via  DP mechanism. Consequently,
there is an increasing spin R/D time background with rising
temperature.

In conclusion, we have performed time-resolved Kerr rotation
measurements on a high-mobility low-density two dimensional electron
gas at low temperatures. We observe that as temperature is
increased, the spin R/D time exhibits a peak of 3.12\ ns around 14\
K, superimposed on an increasing background from 1.03\ ns at 1.5\ K
to 2.67\ ns at 30\ K. The appearance of the peak is ascribed to the
electron-electron Coulomb scattering. As temperature approaches the
point where the crossover from the degenerate to the nondegenerate
regime occurs, the electron-electron scattering becomes strongest.
This results in a peak in spin R/D time versus temperature curve due
to the DP mechanism. Our results nail down the importance of the
Coulomb scattering to the spin R/D due to the DP mechanism.

We thank H. Z. Zheng,  K. Chang and M. W. Wu for fruitful
discussions and M. Heiblum for valuable help. This work has been
supported by NSFC under grant Nos. 10425149 and 10334040 and the
Knowledge Innovation Project of Chinese Academy of Sciences.


\newpage

\end{document}